\documentclass[12pt]{article}
\usepackage{a4wide,epsfig,psfrag,amsmath,amssymb,cite,scalefnt}

\newcommand{\gsim}{\;\rlap{\lower 3.5 pt \hbox{$\mathchar \sim$}} \raise 1pt
 \hbox {$>$}\;}
\newcommand{\lsim}{\;\rlap{\lower 3.5 pt \hbox{$\mathchar \sim$}} \raise 1pt
 \hbox {$<$}\;}

\newcommand{\heavi}[1]{\theta\left(#1\right)}

\begin{document}

\title{\vskip-3cm{\baselineskip14pt
    \begin{flushleft}
      \normalsize SFB/CPP-13-54\\
      \normalsize TTP13-28
  \end{flushleft}}
  \vskip1.5cm
  Colour octet potential to three loops
}
\author{
  Chihaya Anzai$^{(a)}$, 
  Mario Prausa$^{(a)}$, 
  Alexander V. Smirnov$^{(b)}$, 
  \\
  Vladimir A. Smirnov$^{(c)}$, 
  Matthias Steinhauser$^{(a)}$\\[1em]
  {\small\it (a) Institut f{\"u}r Theoretische Teilchenphysik,}\\
  {\small\it Karlsruhe Institute of Technology (KIT)}\\
  {\small\it 76128 Karlsruhe, Germany}
  \\
  {\small\it (b) Scientific Research Computing Center,
    Moscow State University}\\
  {\small\it 119992 Moscow, Russia}
  \\
  {\small\it (c) Nuclear Physics Institute,
    Moscow State University}\\
  {\small\it 119992 Moscow, Russia}\\
}

\date{}

\maketitle

\thispagestyle{empty}

\begin{abstract}

We consider the interaction between two static sources in the 
colour octet configuration and compute the potential to three
loops. Special emphasis is put on the treatment of pinch contributions
and two methods are applied to reduce their evaluation to diagrams
without pinches.

\medskip


\end{abstract}

\thispagestyle{empty}


\newpage


\section{Introduction}

The potential energy between two heavy quarks is a fundamental quantity in
physics. In fact, the history of computing loop corrections to the
potential of quarks forming a colour singlet configuration goes back to the
mid-seventies with the idea to describe a bound state of heavy coloured
objects in analogy to the hydrogen atom~\cite{Appelquist:1974zd}.  One-loop
corrections were computed shortly afterwards in
Refs.~\cite{Fischler:1977yf,Billoire:1979ih}. The two-loop
corrections have only been evaluated towards the end of the nineties by two
groups~\cite{Peter:1996ig,Peter:1997me,Schroder:1998vy} and about five years ago
the three-loop corrections have been considered in
Refs.~\cite{Smirnov:2008pn,Smirnov:2009fh,Anzai:2009tm}, again in
two independent calculations.

In this paper we consider the potential in momentum space which we define as
\begin{eqnarray}
  V^{[c]}(|{\vec q}\,|)&=&
  -{4\pi C^{[c]} \frac{\alpha_s(|{\vec q}\,|)}{{\vec q}\,^2}}
  \Bigg[1+\frac{ \alpha_s(|{\vec q}\,|) }{4\pi} a_1^{[c]}
    +\left( \frac{\alpha_s(|{\vec q}\,|)}{4\pi}\right)^2a_2^{[c]}
    \nonumber\\&&\mbox{}
    +\left(\frac{\alpha_s(|{\vec q}\,|)}{4\pi}\right)^3
    \left(a_3^{[c]}+ 8\pi^2 C_A^3\ln\frac{\mu^2}{{\vec q}\,^2}\right)
    +\cdots\Bigg]\,,
  \label{eq::V}
\end{eqnarray}
where $C_A=N_c$ and $C_F=(N_c^2-1)/(2N_c)$
are the eigenvalues of the quadratic Casimir
operators of the adjoint and fundamental representations of the
$SU(N_c)$ colour gauge group, respectively.
The strong coupling $\alpha_s$ is defined in the $\overline{\rm MS}$ scheme
and for the renor\-mali\-zation scale we choose $\mu=|{\vec q}\,|$
in order to suppress the corresponding logarithms. The general results, both in
momentum and coordinate space, can be found in Appendix~\ref{app::V}.

In Eq.~(\ref{eq::V}) we have introduced the superscript ${[c]}$ which
indicates the colour state of the quark-anti-quark system.  In
Refs.~\cite{Peter:1996ig,Peter:1997me,Schroder:1998vy,Smirnov:2008pn,Smirnov:2009fh,Anzai:2009tm}
only the singlet configurations ($c=1$) have been considered, which is
phenomenologically most important. However, quarks in the fundamental
representation can also combine to a colour octet state.  At tree-level and
one-loop order only the overall colour factor changes from $C^{[1]}=C_F$ to
$C^{[8]}=C_F-C_A/2$. Starting from two
loops~\cite{Kniehl:2004rk,Collet:2011kq} the coefficients
$a_i^{[c]}$ get
additional contributions. In this paper we compute $a_3^{[8]}$ and compare the
result to $a_3^{[1]}$~\cite{Smirnov:2008pn,Smirnov:2009fh,Anzai:2009tm}.

The term proportion to $\ln\mu^2$ in Eq.~(\ref{eq::V}) has its origin
in an infra-red divergence which has been subtracted minimally. It
appears for the first time at three-loop
order~\cite{Appelquist:1977es} and is canceled against the
ultraviolet divergence of the ultrasoft contributions which have been
studied in Refs.~\cite{Brambilla:1999qa,Kniehl:1999ud,Kniehl:2002br}.
As anticipated in Eq.~(\ref{eq::V}), the ultrasoft contribution for
the colour-singlet and colour-octet case differs only by the overall
colour factor which is confirmed by our explicit calculation.

Let us for completeness mention that 
it is possible to generalize the concept of the heavy-quark potential to
generic colour sources which in principle can also be in the adjoint
representation of $SU(3)$ as, e.g., the gluino in supersymmetric theories. 
Various combinations of quark, squark and gluino bound state systems have been
considered in Ref.~\cite{Collet:2011kq} and the corresponding potential has
been evaluated up to two loops.

A further generalization of the three-loop corrections to $V^{[1]}$ 
has been considered in
Ref.~\cite{Anzai:2010td} where it is still assumed that the heavy sources
form a colour singlet state, however, the colour representation is kept general.

The remainder of the paper is organized as follows: in the next Section we
explain in detail how we treat the diagrams involving pinches. Afterwards we
present our results in Section~\ref{sec::results} and conclude
in Section~\ref{sec::conclusions}.


\section{Calculation}

As compared to the singlet case
the calculation of the octet potential is substantially more complicated which
is connected to the occurrence of so-called pinch contributions as shall be
discussed in the following. Pinch contributions occur in those cases where a
deformation of the integration contour, needed to circumvent poles in the 
complex plane of the zero-component of the integration momentum, is not
possible. 

\begin{figure}[t]
  \centering
  \begin{center}
    \begin{tabular}{cccc}
      \includegraphics[width=.2\textwidth]{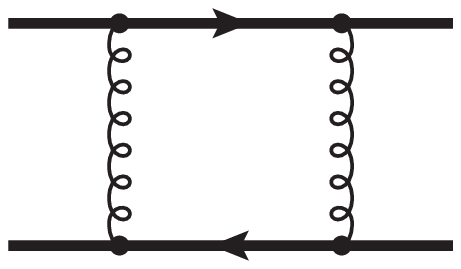} &
      \includegraphics[width=.2\textwidth]{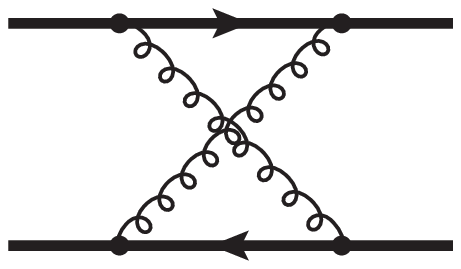} &
      \includegraphics[width=.2\textwidth]{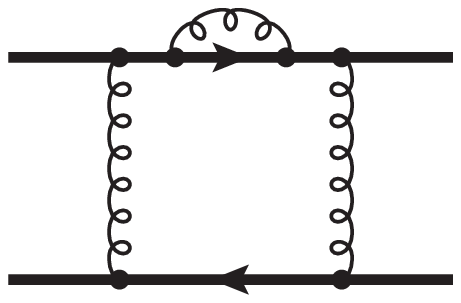} &
      \includegraphics[width=.2\textwidth]{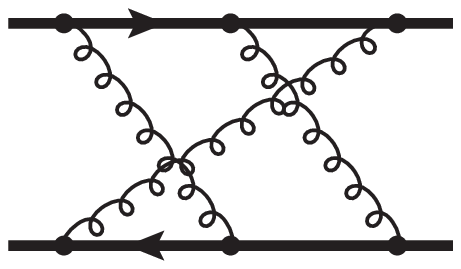} \\
      (a) & (b) & (c) & (d) \\
      \includegraphics[width=.2\textwidth]{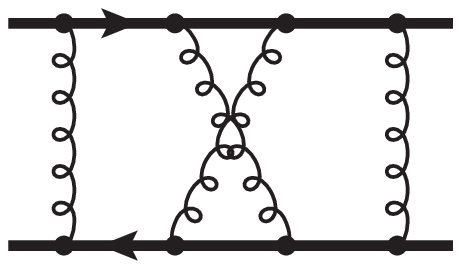} &
      \includegraphics[width=.2\textwidth]{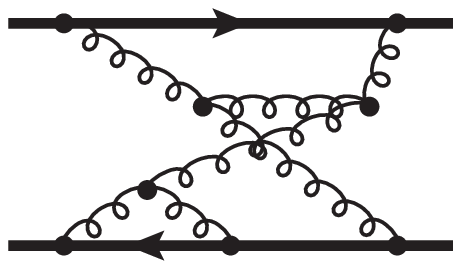} &
      \includegraphics[width=.2\textwidth]{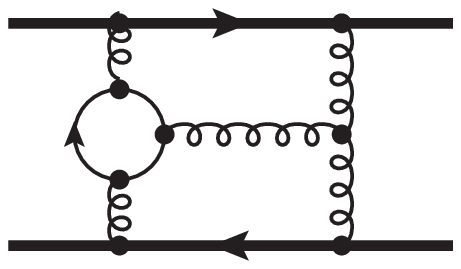} &
      \includegraphics[width=.2\textwidth]{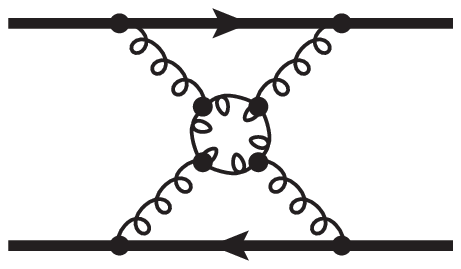} \\
      (e) & (f) & (g) & (h)
    \end{tabular}
  \end{center}
  \caption{\label{fig::diags}Feynman diagrams up to three-loop order
    contributing to $V^{[c]}$. Thick lines represent static quarks,
    thin solid lines massless fermions and curled lines gluons.}
\end{figure}

For illustration let us consider the planar ladder diagram in
Fig.~\ref{fig::diags}(a). Since the momentum transfer $q$ between the heavy
quarks is space-like and the static propagators only contain the 
energy component of the momentum
we obtain for the loop integral the expression
\begin{eqnarray}
  \int {\rm d}^D k \, f(k,\vec{q}\,) \, \frac{1}{(k_0 + i0)(k_0 - i0)}
  \,,
  \label{eq::pinch}
\end{eqnarray}
where $f(\vec{q}\,)$ collects all prefactors and the contribution from the
gluon propagators, and $D=4-2\epsilon$ is the space-time dimension.

There are several possibilities to treat the one-loop diagram in
Eq.~(\ref{eq::pinch}) and obtain a relation to a well-defined
integral.
For example, it is possible to apply the principle value prescription
\begin{eqnarray}
  \frac{1}{(k_0 + i0)(k_0 - i0)}
  &\to&
  \frac{1}{2}\left[{1\over(k_0+i\varepsilon)^{2}}
    +\frac{1}{(k_0-i\varepsilon)^{2}}\right]
  \,,
  \label{eq::pv}
\end{eqnarray}
which is valid in the soft region for the integration momenta.  With the help
of Eq.~(\ref{eq::pv}) it is possible to treat all contributions involving
pinches in one loop
momentum~\cite{Kniehl:2004rk,Collet:2011kq,Prausa:2013qva}.  The application
to diagrams with two or more pinch contributions in one diagram, a
situation which appears at two loops and beyond, is not obvious.

Another possibility is based on the fact that in QED with only one (heavy)
lepton pair the potential between the fermion and anti-fermion is
given by the tree-level term (see, e.g., discussion in
Refs.~\cite{Schroder:1999sg,Smirnov:2008pn}) which means that the loop
corrections are exactly canceled by the iteration terms of lower-order
contributions. The latter arise from the fact that the potential is
proportional to the logarithm of the quark-anti-quark four-point
amplitude which has to be expanded in the coupling constant.
Translating this knowledge to QCD means that the sum of all one-loop
contributions proportional to $C_F^2$ have to vanish which in turn
leads to the graphical equation\footnote{We denote the static quarks
  by horizontal thick lines and the massless modes by thin
  lines connecting the colour sources.}
\begin{eqnarray}
  \begin{matrix}
    \includegraphics{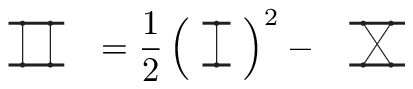} \,.
  \end{matrix}
  \label{eq::1lbox}
\end{eqnarray}
In this way the planar-ladder contribution can be replaced by the
crossed ladder which is free of pinch contributions.
The same method has successfully been applied at two
loops~\cite{Kniehl:2004rk,Collet:2011kq,Prausa:2013qva} [see
Ref.~\cite{Prausa:2013qva} for the two-loop analogue of Eq.~(\ref{eq::1lbox})].

In the following we provide a general prescription for the treatment of the
pinch contributions which works 
to all loop orders and for arbitrary number of involved propagators.
We formulate the algorithm in a way which is convenient
for our application. Alternative formulations can be found in
Refs.~\cite{Fischler:1977yf,Schroder:1999sg}.

\begin{figure}[t]
  \centering
  \includegraphics[width=.4\textwidth]{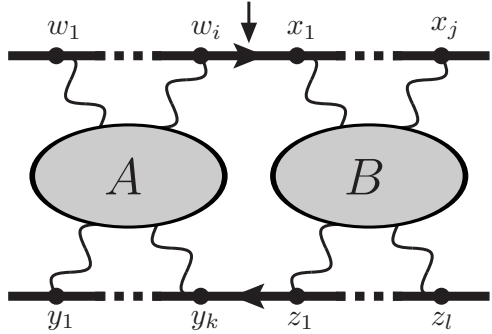}
  \caption{\label{fig::diag_gen}Generic  Feynman diagram
    involving a pinch contribution. The blobs $A$ and $B$ may contain
    further pinches which are treated recursively (see description of
    algorithm). The arrow indicates where the static line is
    cut at first.
    }
\end{figure}

It is convenient to formulate the algorithm in coordinate space. Using the
Feynman rules from Appendix~\ref{app::FR} the one-loop ladder diagram in
Fig.~\ref{fig::diags}(a) takes the form 
\begin{eqnarray}
    &&g_s^4 \int\limits_{-\frac T2}^{\frac T2} dw_0 \int\limits_{-\frac
    T2}^{\frac T2} dx_0 \int\limits_{-\frac T2}^{\frac T2} dy_0
  \int\limits_{-\frac T2}^{\frac T2} dz_0\; \heavi{x_0 - w_0}
  \heavi{z_0 - y_0} \nonumber\\ &&\hspace{25ex} \times
  D_{00}\left(w_0-y_0,\vec{r}\right)
  D_{00}\left(x_0-z_0,\vec{r}\right)\,,
  \label{eq::ladder}
\end{eqnarray}
where the colour structure has been ignored. In Eq.~(\ref{eq::ladder})
the integration over the spacial components has been performed and the
$\theta$ functions of the vertices have been used to restrict the integration
limits of the temporal integrals to $[-T/2,T/2]$.
Note that $\theta(x_0-w_0)$ refers to the upper and 
$\theta(y_0-z_0)$ to the lower source line.
In analogy one obtains for the generic diagram in Fig.~\ref{fig::diag_gen}
the following combination of $\theta$ functions:
\begin{eqnarray}
  \heavi{w^0_2-w^0_1}\dots\heavi{w^0_i-w^0_{i-1}}
  \heavi{x^0_1-w^0_i}\heavi{x^0_2-x^0_1}\dots\heavi{x^0_j-x^0_{j-1}}
  \,.
  \label{eq::theta}
\end{eqnarray}

A crucial ingredient to the algorithm described below is the cut of a static
propagator. This is equivalent to setting the corresponding propagator to
unity, i.e., the associated $\theta$ function in Eq.~(\ref{eq::theta}) is
set to one.  Actually, the omission of $\theta(x_1-w_i)$ from
Eq.~(\ref{eq::theta}) (see also Fig.~\ref{fig::diag_gen}) relaxes the original
conditions on the zero components
\begin{eqnarray}
  w^0_1 < \dots < w^0_i < x^0_1 < \dots < x^0_j\text. 
  \label{eq::cond1}
\end{eqnarray}
to
\begin{eqnarray}
  w^0_1 < \dots < w^0_i \wedge x^0_1 < \dots < x^0_j
  \,.
  \label{eq::cond2}
\end{eqnarray}
The latter is satisfied by all Feynman diagrams which are obtained from the
original one by permutations of vertices in the upper source line as long as
the order of the vertices involving $w$'s and $x$'s is kept. Thus the result
for the cut diagram is obtained by summing all such contributions.  Let us
illustrate this mechanism by the following three-loop diagram
\begin{eqnarray}
  \begin{matrix}
    \includegraphics{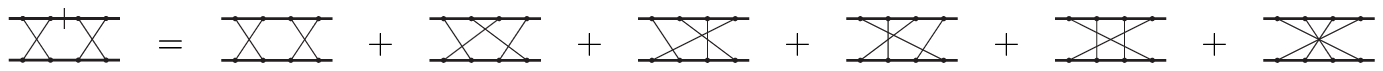} \,.
  \end{matrix}
  \label{eq::cut}
\end{eqnarray}
The procedure for cutting an anti-source propagator is, of course, in close
analogy.

We are now in the position to describe the algorithm which can be
applied to all diagrams involving pinch contributions. The output of
the algorithm are equations which relate pinch diagrams to diagrams
without pinches. The latter can be computed along the lines of
Refs.~\cite{Smirnov:2008pn,Smirnov:2009fh,Anzai:2009tm}.  In all steps
QED-like colour factors are assumed; the multiplication with the
proper colour factor happens after applying the obtained relations.
In parallel to the description of the algorithm we illustrate its
principle of operation on explicit two- and three-loop examples.

\begin{enumerate}

\item \label{setp::1} Consider a diagram with a pinch. In case there is more
  than one pinch the following steps have to be applied to each one
  consecutively.  In case more than one source or anti-source propagator is
  involved [see, e.g., Fig.~\ref{fig::diags}(c)] the operations are performed
  for the most left and most right propagator and the resulting equations are
  added.

\item Express the pinch diagram by the corresponding diagram with a cut source
  propagator and the remaining contributions according to Eq.~(\ref{eq::cut}).

  For our three-loop example the corresponding equation looks as
  follows 
  \begin{eqnarray}
    \begin{matrix}
      \includegraphics{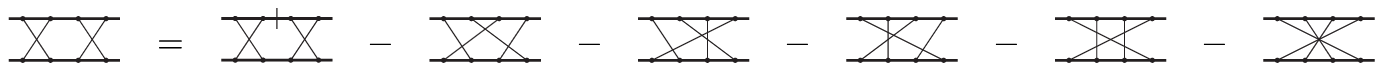} \,.
    \end{matrix}
    \label{eq::ex_1}
  \end{eqnarray}

\item Replace the diagram with a cut source propagator by the diagram where
  both the source and anti-source propagators are cut and the remaining
  contributions which are obtained in analogy to step~\ref{setp::1}. Write
  the diagram with two cuts as a product of lower-order contributions.

  The application of these rules to our example leads to
  \begin{eqnarray}
    \begin{matrix}
      \includegraphics{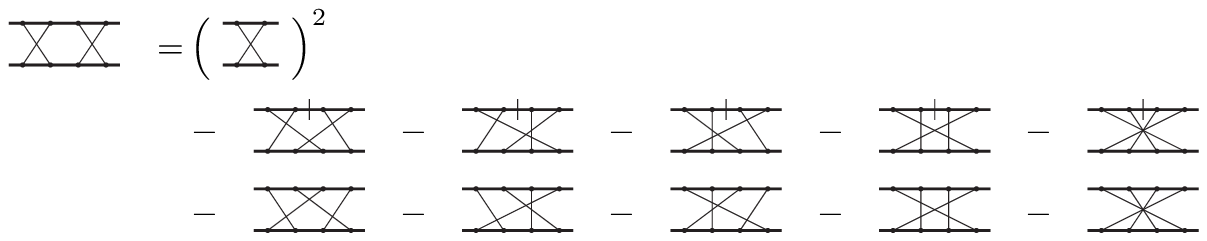} \,.
    \end{matrix}
    \label{eq::ex_2}
  \end{eqnarray}

\item In a next step the diagrams with a cut in the source propagator have to
  be treated. This is done by replacing them by the sum of diagrams obtained
  by considering all allowed permutations of the source vertices.

  In our example this leads to
  \begin{eqnarray}
    \begin{matrix}
      \includegraphics{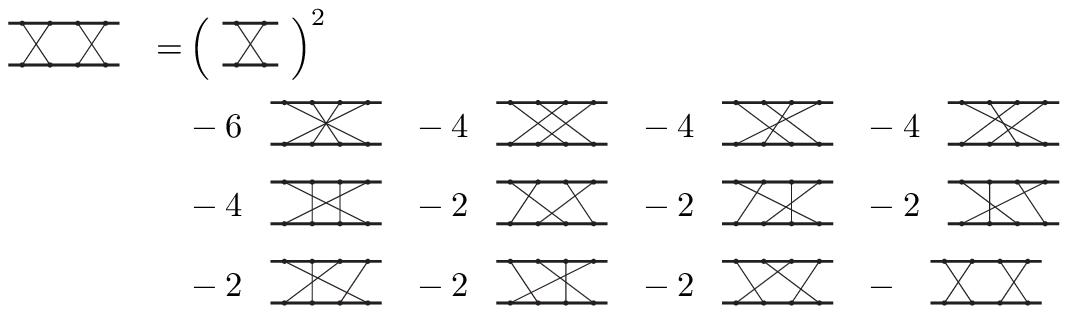} \,.
    \end{matrix}
    \label{eq::ex_3}
  \end{eqnarray}

\item\label{step_sol} Solve the resulting equation for the considered
  pinch diagram. 

  In our example the original diagram appears on the right-hand side
  with a negative sign. This finally leads to
  \begin{eqnarray}
    \begin{matrix}
      \includegraphics{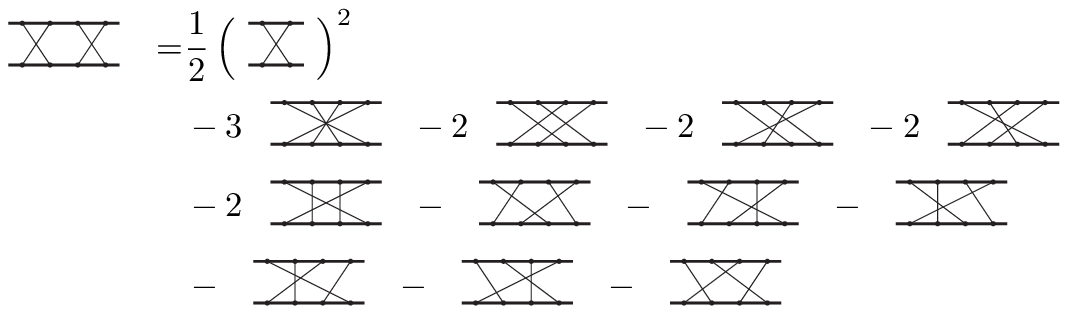} \,.
    \end{matrix}
    \label{eq::ex_4}
  \end{eqnarray}
  Note that all diagrams on the right-hand side
  of this equation are either products of lower-order contributions or 
  are free of pinches and can thus be computed in the standard way.

\item It might be that during the described procedure scaleless
  integrals appear which are set to zero within dimensional
  regularization. This is in particular true for non-amputated
  diagrams.

\item It is advantageous to add diagrams with the same colour factor
  before applying the described algorithm since in some cases
  the pinch diagram appears on the right-hand side which are symmetric to the 
  original one, however, step~\ref{step_sol} can not be performed.

  As an example consider the two two-loop diagrams which can be written as
  \begin{eqnarray}
    \begin{matrix}
      \includegraphics{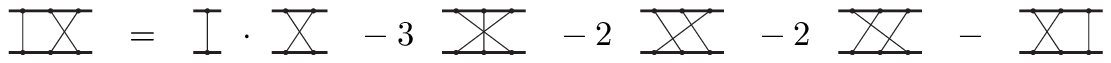} \,, \\
      \includegraphics{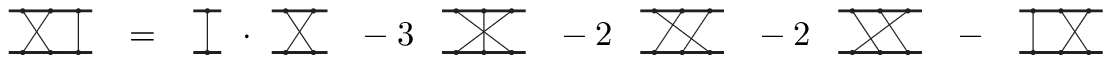} \,.
    \end{matrix}
    \label{eq::ex_5}
  \end{eqnarray}
  In this version the equations can not be used. However, the sum of the equations
  leads to 
  \begin{eqnarray}
    \begin{matrix}
      \includegraphics{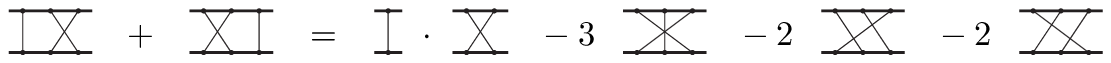} \,.
    \end{matrix}
    \label{eq::ex_6}
  \end{eqnarray}

\item In case there are still pinch contributions on the right-hand side of
  the equation the described procedure is applied iteratively.

  Consider, e.g., the two-loop ladder diagram which, after applying the above
  steps once, leads to 
  \begin{eqnarray}
    \begin{matrix}
      \includegraphics{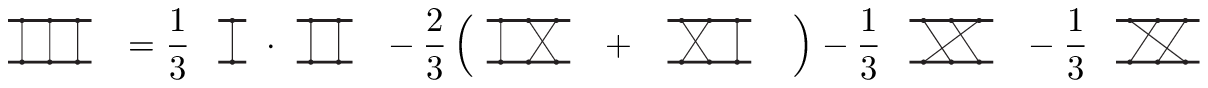} \,.
    \end{matrix}
    \label{eq::ex_7}
  \end{eqnarray}
  Using Eq.~(\ref{eq::ex_6}) and the one-loop relation~(\ref{eq::1lbox})
  results in the equation
  \begin{eqnarray}
    \begin{matrix}
      \includegraphics{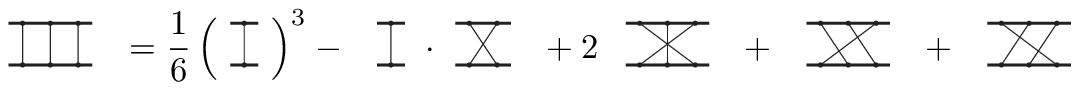} \,.
    \end{matrix}
    \label{eq::ex_8}
  \end{eqnarray}
\end{enumerate}

It is straightforward to implement the described algorithm in a computer
program. We have verified that we reproduce the two-loop results in the
literature~\cite{Kniehl:2004rk,Collet:2011kq,Prausa:2013qva}. 
Furthermore, we have verified the exponentiation of the singlet potential
up to three loops and we have checked that all iteration terms predicted from
lower-order contributions are reproduced.

Sample Feynman diagrams contributing to $V^{[8]}$ up to three loops are
shown in Fig.~\ref{fig::diags}. We generate the amplitudes with the
help of {\tt qgraf}~\cite{Nogueira:1991ex} and process the output
further using {\tt q2e} and {\tt
  exp}~\cite{Harlander:1997zb,Seidensticker:1999bb} in order to arrive
at {\tt FORM}-readable results.  At that point projectors are applied,
traces are taken and the scalar products in the numerator are
decomposed in terms of denominator factors. The resulting scalar
expressions are mapped to the integral families defined in
Refs.~\cite{Smirnov:2008tz,Smirnov:2008ay,Smirnov:2010zc,Smirnov:2010gi}.

The algorithm used for the treatment of the pinch contributions described above 
is applied to the output file of {\tt qgraf}. While adding the results of
all diagrams the obtained relations are applied and thus the sum is expressed
in terms of well-defined integrals.

The computation of the integrals proceeds along the lines of
Refs.~\cite{Smirnov:2008pn,Smirnov:2009fh,Anzai:2009tm}.  In particular
  we use {\tt FIRE}~\cite{Smirnov:2008iw,Smirnov:2013dia} to reduce the
  integrals to a minimal set, the so-called master integrals.  Actually, only
of the order of one hundred integrals remain to be reduced after using the
tables generated for the computation of the three-loop singlet contribution.
They are quite simple and require only a few days of CPU time.

We managed to express the final result for $a_3^{[8]}$ in terms of the
same 41 master integrals as $a_3^{[1]}$. Thus, our final result only
contains three coefficients (of the $\epsilon$ expansion) which are
not yet known analytically.  Details on the computation of the master
integrals can be found in
Refs.~\cite{Smirnov:2008ay,Smirnov:2010zc,Smirnov:2010gi,Anzai:2012xw}.
For most of the integrals even explicit results are provided.  For
example, the 14 master integrals containing a massless one-loop
subdiagram can be found in~\cite{Smirnov:2010gi} and 16 integrals
among the most complicated ones are provided in
Ref.~\cite{Smirnov:2010zc}.

We have performed a second calculation of $a_3^{[8]}$ which is
described in the following. The calculation of the loop integrals
is based on Refs.~\cite{Anzai:2009tm,Anzai:2010td} and for the
treatment of the pinch contributions we follow
Refs.~\cite{Gatheral:1983cz,Frenkel:1984pz}. The basic idea outlined
in these references is that the colour factor of the diagrams without
pinch is changed in such a way that the pinch contribution is taken
into account. At the same time the pinch diagrams are set to zero.

To reach this goal we define for a colour diagram $x$
\begin{eqnarray}
  E\left(x\right) &:=& C\left(x\right) -
  \sum_{d\in\mathrm{Dec'}\left(x\right)}T^{-n\left(d\right)}
  E\left(d\right)\,,
  \label{eq::Ex}
\end{eqnarray}
which corresponds to Eq.~(4) of Ref.~\cite{Gatheral:1983cz}.  In this
equation $\mathrm{Dec}'\left(x\right)$ represents the set of
nontrivial decompositions of $x$ and $n\left(d\right)$ is the number
of webs\footnote{A web is a set of gluons which cannot be partitioned
  without cutting at least one of its lines and a decomposition $d$ of
  a set of gluon lines is a classification of the lines into webs such
  that each line is precisely in one web~\cite{Gatheral:1983cz}.  See
  Ref.~\cite{Gatheral:1983cz} for examples.}  
in $d$.  $T$ equals $N_{c}$ for the singlet case and
$N_{c}C_{F}$ for the octet case.

Each Feynman diagram $F$ can be expressed in terms of a product of the colour
factor $C\left(F\right)$ and the momentum space integral $I\left(F\right)$. If
the colour factor $C\left(F\right)$ of each diagram is replaced by the new
colour factor $E\left(F\right)$ calculated with the help of
Eq.~(\ref{eq::Ex}), all contributions from iterations will be eliminated.
Moreover, $E\left(F\right)$ is zero for all pinch diagrams and
hence their evaluation is not needed anymore.

\newsavebox{\boxao}\sbox{\boxao}{%
\begin{minipage}[c]{0.08\columnwidth}%
\includegraphics[width=1\textwidth]{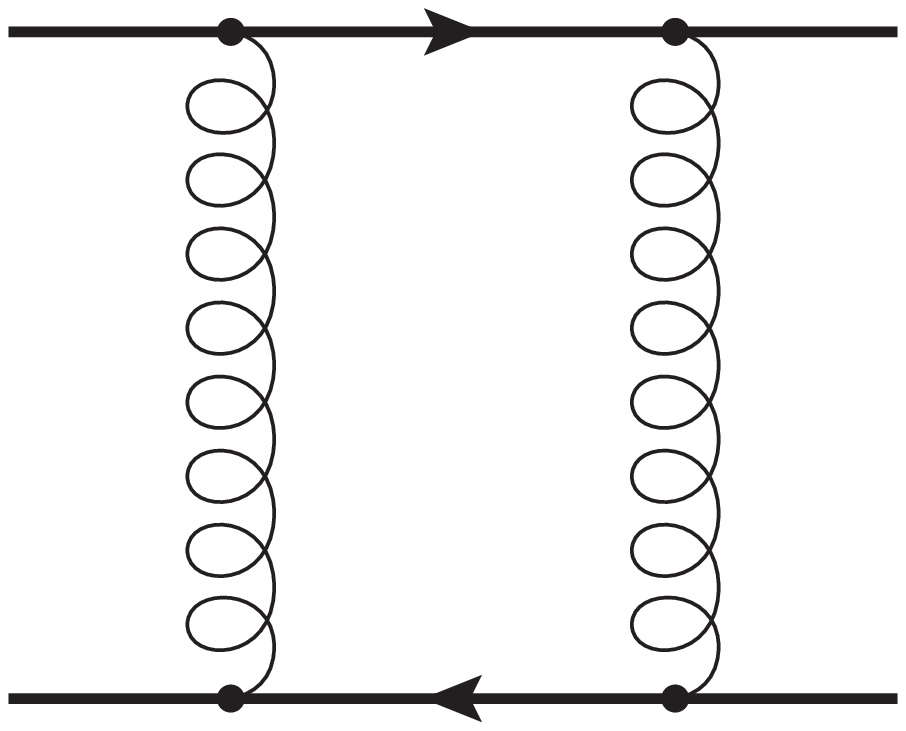}%
\end{minipage}}

\newsavebox{\boxap}\sbox{\boxap}{%
\begin{minipage}[c]{0.08\columnwidth}%
\includegraphics[width=1\textwidth]{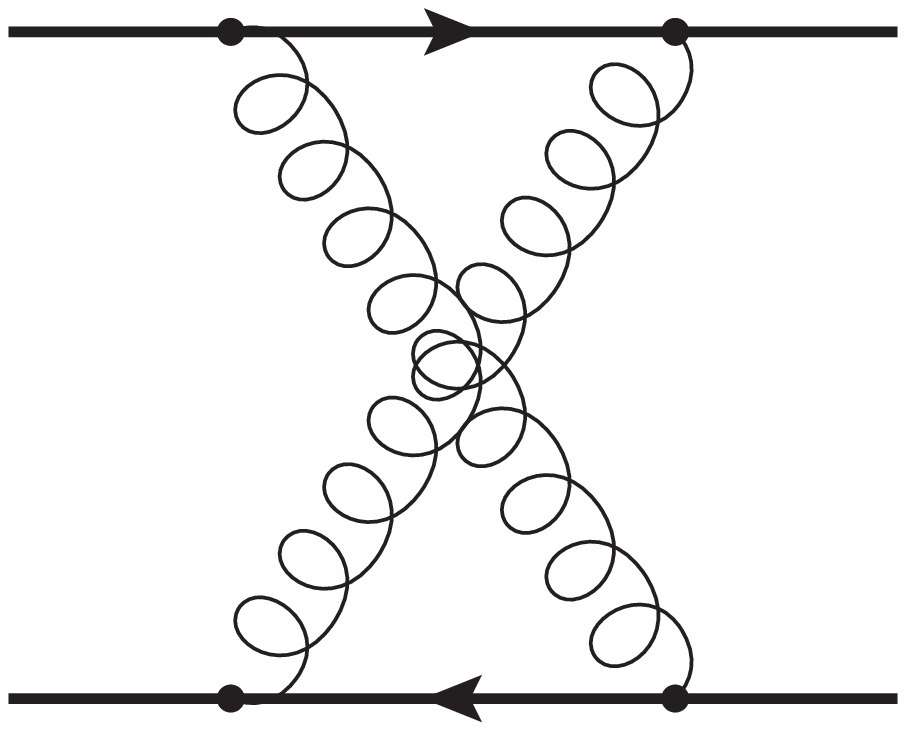}%
\end{minipage}}

\newsavebox{\boxar}\sbox{\boxar}{%
\begin{minipage}[c]{0.06\columnwidth}%
\includegraphics[width=1\textwidth]{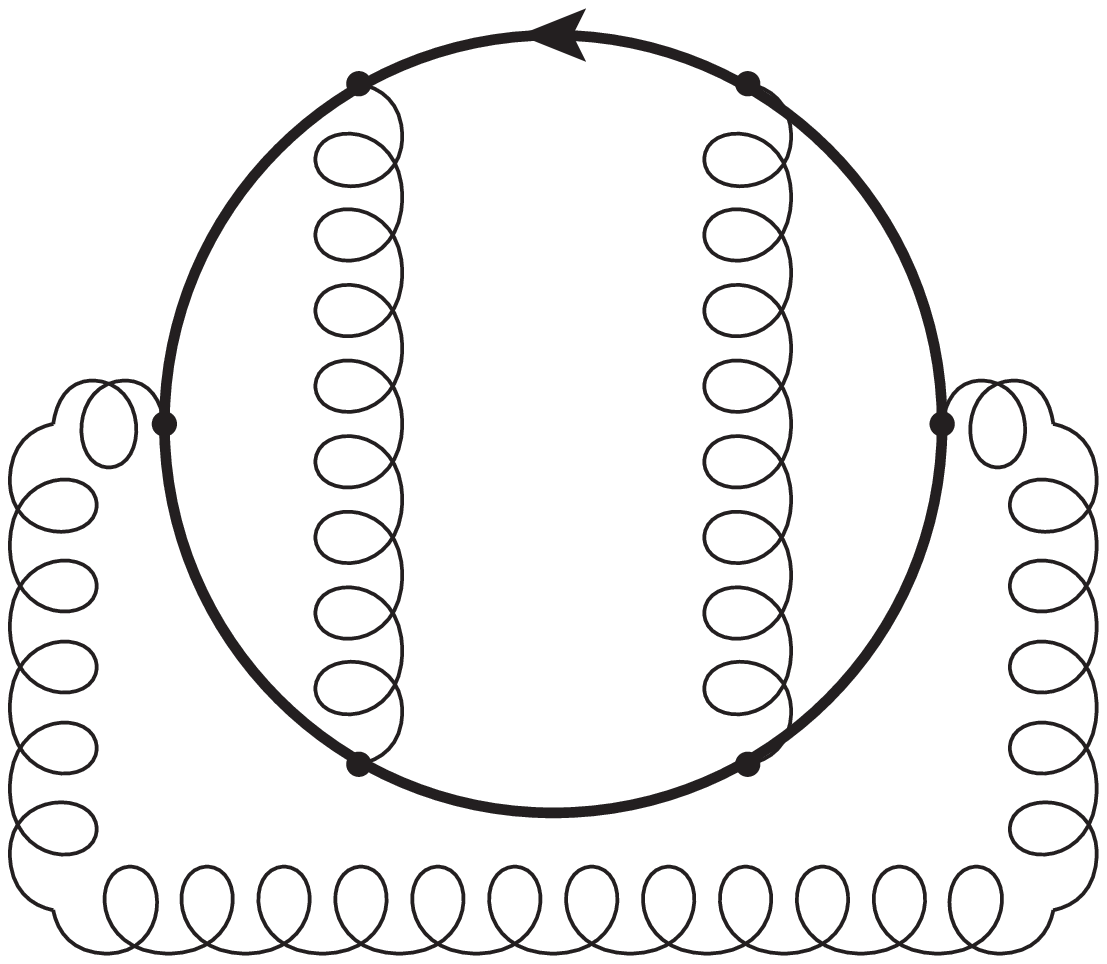}%
\end{minipage}}

\newsavebox{\boxaj}\sbox{\boxaj}{%
\begin{minipage}[c]{0.06\columnwidth}%
\includegraphics[width=1\textwidth]{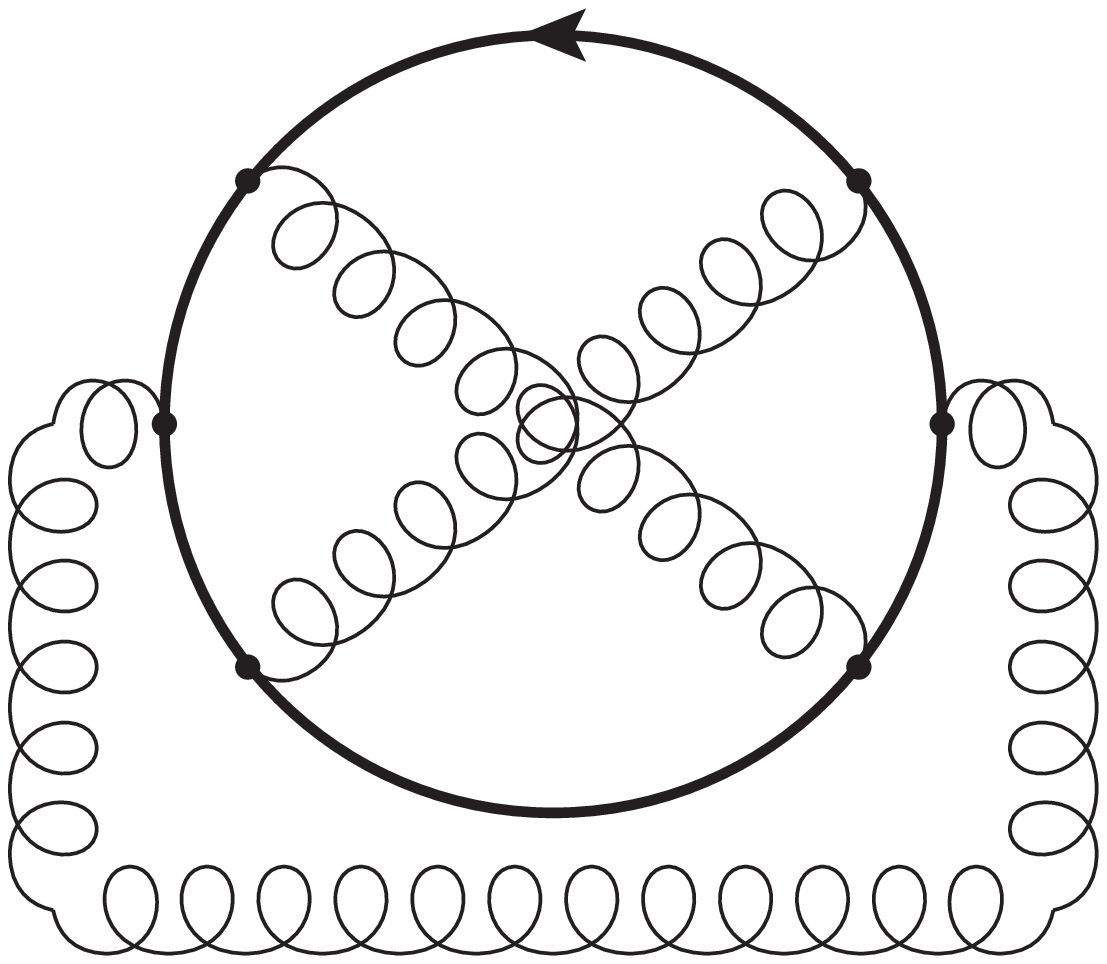}%
\end{minipage}}

\newsavebox{\boxaf}\sbox{\boxaf}{%
\begin{minipage}[c]{0.06\columnwidth}%
\includegraphics[width=1\textwidth]{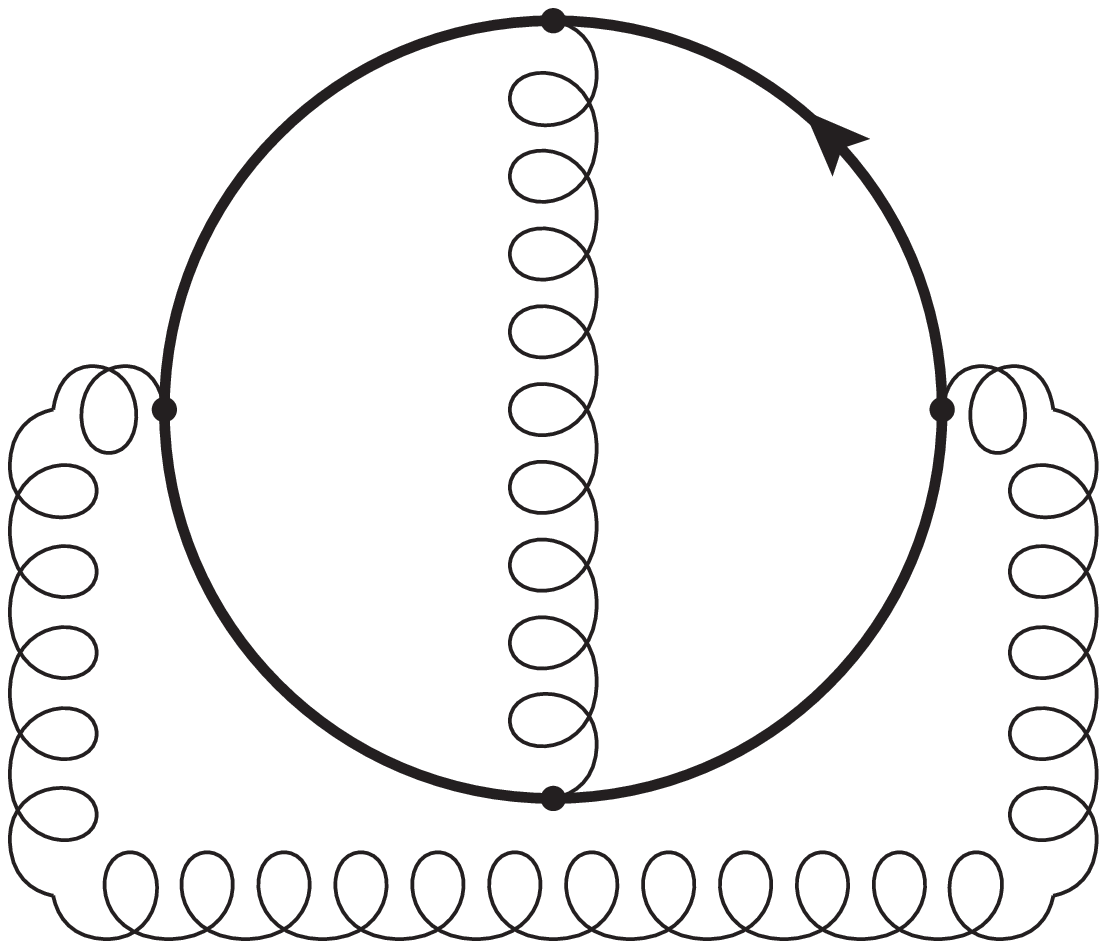}%
\end{minipage}}

Let us for illustration consider the contribution from the 
one-loop planar and crossed ladder diagrams of
Fig.~\ref{fig::diags}(a) and~(b). For the octet potential
we can write
\begin{eqnarray}
  \delta V_{\rm ladder}^{[8],(1)} &=& I\left(\usebox{\boxao}\right) \times
  \frac{1}{N_{c}C_{F}}\usebox{\boxar}
  +I\left(\usebox{\boxap}\right) \times
  \frac{1}{N_{c}C_{F}}\usebox{\boxaj}
  \,,
  \label{eq::with itr}
\end{eqnarray}
where the colour factors $C(F)$ are presented in graphical from
after the factor $1/(N_c C_F)$.
If one now replaces $C(F)$ by $E(F)$ and used Eq.~(\ref{eq::Ex})
one obtains
\begin{align*}
  V^{[8],(1)}_{\rm ladder}|_{C\rightarrow E} &
  =I\left(\usebox{\boxao}\right)\times\frac{1}{N_{c}C_{F}}\left[\usebox{\boxar}-\frac{1}{N_{c}C_{F}}\left(\usebox{\boxaf}\right)^{2}\right]\\ &
  \qquad\qquad+I\left(\usebox{\boxap}\right)\times\frac{1}{N_{c}C_{F}}\left[\usebox{\boxaj}-\frac{1}{N_{c}C_{F}}\left(\usebox{\boxaf}\right)^{2}\right]\\ &
  =I\left(\usebox{\boxao}\right)\times0+I\left(\usebox{\boxap}\right)\times\frac{1}{N_{c}C_{F}}\left[\usebox{\boxaj}-\usebox{\boxar}\right]
  \,,
\end{align*}
which is equivalent to the relation~(\ref{eq::1lbox}) obtained with the method
described above.

In the results which we present below both methods
described in this Section lead to the same final expressions which is a
strong check for their correctness.


\section{\label{sec::results}Results for $V^{[8]}$}

In this section we present results for the coefficients $a_i^{[c]}$ in
Eq.~(\ref{eq::V}) for $SU(N_c)$ with generic number of colours, $N_c$.
Let us for convenience repeat the one- and two-loop results which
read (The octet results have been obtained in
Refs.~\cite{Kniehl:2004rk,Collet:2011kq}.) 
\begin{eqnarray}
  a_1^{[1]} &=& \frac{31}9 C_A - \frac{20}9 T_F n_l \,,\nonumber\\
  a_2^{[1]} &=& \left(\frac{4343}{162} + 4\pi^2 - \frac{\pi^4}4 + \frac{22}3 \zeta(3)\right) C_A^2 
      - \left(\frac{1798}{81} + \frac{56}3 \zeta(3)\right) C_A T_F n_l \nonumber \\
      &&\quad
      - \left(\frac{55}3 - 16\zeta(3)\right) C_F T_F n_l 
      + \left(\frac{20}9\right)^2 T_F^2 n_l^2 \,, \nonumber \\
  a_1^{[8]} &=& a_1^{[1]}\,,\nonumber\\
  a_2^{[8]} &=& a_2^{[1]} + N_c^2 \pi^2 \left(\pi^2-12\right)\,.
  \label{eq::a_12}
\end{eqnarray}
It is remarkable that the difference between the singlet and octet
contribution at two loops involves only $\pi^2$ and $\pi^4$ terms.

At three-loop order it is convenient to decompose the coefficient
in the form
\begin{eqnarray}
  a_3^{[c]} &=& a_3^{[c],(3)} n_l^3 + a_3^{[c],(2)} n_l^2 +  
  a_3^{[c],(1)} n_l +  a_3^{[c],(0)}
  \,,
\end{eqnarray}
where $n_l$ is the number of light quarks. For the 
first two coefficients we have
\begin{eqnarray}
  a_3^{[1],(3)} &=& -\left(\frac{20}9\right)^3 T_F^3 \,,\nonumber\\
  a_3^{[1],(2)} &=& \left(\frac{12541}{243} + \frac{368\zeta(3)}3 + \frac{64\pi^4}{135}\right) C_A T_F^2
      + \left(\frac{14002}{81} - \frac{416\zeta(3)}3\right) C_F T_F^2 \,,\nonumber\\
  a_3^{[8],(3)} &=& a_3^{[1],(3)}\,,\nonumber\\
  a_3^{[8],(2)} &=& a_3^{[1],(2)}\,.
  \label{eq::a3a}
\end{eqnarray}
For the coefficients $a_3^{[c],(1)}$ and $a_3^{[c],(0)}$
we expect a similar feature as in the two-loop result of 
Eq.~(\ref{eq::a_12}) and thus we write
\begin{eqnarray}
  a_3^{[1],(1)} &=& -709.717 \, C_A^2 T_F +
  \left(-\frac{71281}{162} + 264\zeta(3) + 80\zeta(5)\right) C_A C_F
  T_F \nonumber\\ &&\quad + \left(\frac{286}9 + \frac{296\zeta(3)}3 -
  160\zeta(5)\right) C_F^2 T_F - 56.83(1) \,
  \frac{d_F^{abcd} d_F^{abcd}}{N_A}
  \nonumber\\ &=& 
  - 367.319 \, N_c^2 
  + 17.3611(7) 
  - 12.597(2)\, \frac{1}{N_c^2}\,, 
  \nonumber\\ 
  a_3^{[8],(1)} &=&
  a_3^{[1],(1)} + \delta a_3^{[8],(1)} 
  \,,\nonumber\\ 
  a_3^{[1],(0)}
  &=& 502.24(1) \, C_A^3  -136.39(12)\,
  \frac{d_F^{abcd} d_A^{abcd}}{N_A} 
  \nonumber\\ &=& 
  - 17.049(7) \, N_c 
  + 499.396   \, N_c^3 \,,
  \nonumber\\ 
  a_3^{[8],(0)} &=&
  a_3^{[1],(0)} + \delta a_3^{[8],(0)} \,,
  \label{eq::a3b}
\end{eqnarray}
with
\begin{eqnarray}
  \delta a_3^{[8],(1)} &=& 6.836(1) + 40.125 \, N_c^2 \,,\nonumber\\
  \delta a_3^{[8],(0)} &=& -97.579(16) \, N_c^3 \,.
  \label{eq::delta_a3_8}
\end{eqnarray}

By comparing Eq.~(\ref{eq::delta_a3_8}) with the 
results in Eq.~(\ref{eq::a3b}) expressed in terms of $N_c$
one observes that the coefficients 
of $n_l/N_c^2$ and $N_c$ are identical for the singlet and octet case
and differences only occur in $n_l N_c^2$, the $N_c$-independent $n_l$ term,
and the $N_c^3$ contribution.
In this context it is interesting to present the complete result for
$a_3^{[8],(1)}$ which reads
\begin{eqnarray}
  a_3^{[8],(1)} &=& -327.193 \, N_c^2 
  + \frac{66133}{648}
  - \frac{112\pi^2}{9}
  - \frac{272\zeta(3)}{3}
  + \frac{8\pi^4}{3}
  - \frac{32\pi^2\zeta(3)}{3}
  + 20\zeta(5)
  \nonumber\\&&\mbox{}
  - 12.597(2)\, \frac{1}{N_c^2}
  \,.
  \label{eq::a3_tot}
\end{eqnarray}
In contrast to the singlet case in Eq.~(\ref{eq::a3b}) it is possible
to obtain an analytic result for the $N_c$-independent part.

Unfortunately, the quantities $\delta a_3^{[8],(0)}$ and $\delta
a_3^{[8],(1)}$ are only available numerically. Thus, it is not immediately
possible to check the analytic structure of the difference between the singlet
and octet coefficient. Nevertheless it is possible to show that
it contains a factor $\pi^2$, a feature which is
also observed at two-loop order~\cite{Kniehl:2004rk,Collet:2011kq} and
for ${\cal N}=4$ supersymmetric Yang Mills theories~\cite{Prausa:2013qva}.
The proof of this claim is based on the observation that
the master integrals which are present in the expressions for
$\delta a_3^{[8],(1)}$ and $\delta a_3^{[8],(1)}$ are of the form
\begin{eqnarray}
  I &=& \int\int\int 
  \frac{{\rm d}^Dk}{(4\pi)^D} \frac{{\rm d}^Dp}{(4\pi)^D} 
  \frac{{\rm d}^Dl}{(4\pi)^D}
  \frac{1}{k_0 + i0}\frac{1}{p_0 + i0} \, f(k,p,l,q)
  \,,
\end{eqnarray}
where $q$ is the external momentum. The integrand has the special property
that one can find variable transformations of $k$, $p$ and $l$ 
which leave the form invariant except for the static propagators in front of
$f(k,p,l,q)$. In fact, one can show that the following relations hold
\begin{eqnarray}
  I &=& \int\int\int 
  \frac{{\rm d}^Dk}{(4\pi)^D} \frac{{\rm d}^Dp}{(4\pi)^D} 
  \frac{{\rm d}^Dl}{(4\pi)^D}
  \frac{1}{-k_0 + i0}\frac{1}{p_0 + i0} \, f(k,p,l,q)
  \nonumber\\
  &=& \int\int\int 
  \frac{{\rm d}^Dk}{(4\pi)^D} \frac{{\rm d}^Dp}{(4\pi)^D} 
  \frac{{\rm d}^Dl}{(4\pi)^D}
  \frac{1}{k_0 + i0}\frac{1}{-p_0 + i0} \, f(k,p,l,q)
  \nonumber\\
  &=& \int\int\int 
  \frac{{\rm d}^Dk}{(4\pi)^D} \frac{{\rm d}^Dp}{(4\pi)^D} 
  \frac{{\rm d}^Dl}{(4\pi)^D}
  \frac{1}{-k_0 + i0}\frac{1}{-p_0 + i0} \, f(k,p,l,q)
  \,.
\end{eqnarray}
Adding the four representations of $I$ leads to
\begin{eqnarray}
  I &=& \frac{1}{4} \int\int\int 
  \frac{{\rm d}^Dk}{(4\pi)^D} \frac{{\rm d}^Dp}{(4\pi)^D} 
  \frac{{\rm d}^Dl}{(4\pi)^D}
  \left( \frac{1}{k_0 + i0} + \frac{1}{-k_0 + i0} \right)
  \left( \frac{1}{p_0 + i0} + \frac{1}{-p_0 + i0} \right)
  \nonumber\\&&\mbox{} 
  \times f(k,p,l,q)
  \,.
\end{eqnarray}
The expressions in the round brackets can be identified with 
$(-2\pi i)\delta(k_0)$ and $(-2\pi i)\delta(p_0)$, respectively,
which immediately leads to an overall factor $\pi^2$.

\begin{table}[t]
  \begin{center}
    \begin{tabular}{c||r|r|r||r|r|r||r|r|r}
      & \multicolumn{3}{c||}{$a_1^{[c]}/4$}
      & \multicolumn{3}{c||}{$a_2^{[c]}/4^2$}
      & \multicolumn{3}{c}{$a_3^{[c]}/4^3$}
      \\
      \hline
      $n_l$ & 3 & 4 & 5 & 3 & 4 & 5 & 3 & 4 & 5\\
      \hline
      singlet 
      & 1.750 & 1.472  & 1.194
      & 16.80 & 13.19  & 9.740
      & 81.25 & 49.39  & 22.83 \\
      \hline
      octet 
      & 1.750  & 1.472   & 1.194
      & 4.973  &  1.366  & $-$2.087
      & 57.33  &  31.22  & 10.41
    \end{tabular}
    \caption{\label{tab::num}Numerical values for the coefficients 
    of $[\alpha_s(\mu=|\vec{q}\,|)/\pi]^i$ ($i=1,2,3$) of the singlet and
    octet potential.} 
  \end{center}
\end{table}

In Tab.~\ref{tab::num} we present numerical results for the coefficients of
$(\alpha_s/\pi)^i$ ($i=1,2,3$) both for the singlet and the octet potential
where for the number of light quarks, $n_l$, we choose the values $3, 4$ and
$5$, which corresponds to the charm, bottom and top quark case, and for the
renormalization scale $\mu=|\vec{q}\,|$. At two-loop order one observes a
compensation of the relatively large two-loop singlet contribution by the
additional term present in the octet case. This term is $n_l$ independent
which even leads to negative values for $a_2^{[8]}$ for $n_l=5$.  Also at
three loops the additional term is negative for all considered values of $n_l$
and leads to a significant reduction, for $n_l=5$ by more than a factor two.


\section{\label{sec::conclusions}Conclusions}

In this paper we have have computed the potential between two heavy
quarks is a colour-octet configuration to three-loop order.
The computation of the underlying integrals profits from the 
calculation of the singlet potential performed in
Refs.~\cite{Smirnov:2008pn,Smirnov:2009fh,Anzai:2009tm}, However, in
contrast to the singlet case the octet potential receives
contributions form diagrams with pinches which significantly
complicates the calculation. We discussed two algorithms
which are used to obtain the pinch contributions by reducing the 
calculation to integrals without pinches.

Our final result is presented in
Eqs.~(\ref{eq::a3a}),~(\ref{eq::a3b}),~(\ref{eq::delta_a3_8})
and~(\ref{eq::a3_tot}). One observes quite some similarity to the  
singlet result. Actually, expressing the coefficients $a_3^{[1]}$ and
$a_3^{[8]}$ in terms of $N_c$ we observe that two out of five
coefficients are identical.

As a physical application of the octet potential 
one can think of top quarks produced at hadron colliders in a
colour-octet state. For the description of the
threshold effects the octet potential serves as a crucial
ingredient (see, e.g., Refs~\cite{Hagiwara:2008df,Kiyo:2008bv}).
Note, however, that the precision of the current calculations
does not yet require three-loop corrections to the potential.
In a further possible application one could use $V^{[8]}$ 
in order to compare with lattice simulations of the potential.


\section*{Acknowledgements}

We would like to thank Alexander Penin for 
carefully reading the manuscript and for useful comments.
This work was supported by DFG through SFB/TR~9.
The work of A.S. and V.S. was partly supported by the Russian Foundation
for Basic Research through grant 11-02-01196.


\begin{appendix}


\section{\label{app::V}$V^{[c]}$ in coordinate and momentum space for
  general renormalization scale $\mu$}

In coordinate space Eq.~(\ref{eq::V}) generalized to arbitrary values of the
renormalization scale reads
\begin{eqnarray}
  \tilde V^{[c]}&=&
  -{C^{[c]} \alpha_s(\mu)\over{r}}
  \Bigg[1+{\alpha_s(\mu)\over 4\pi}\tilde c_1^{[c]}(\mu r)
    +\left({\alpha_s(\mu)\over 4\pi}\right)^2\tilde c_2^{[c]}(\mu r)
    \nonumber\\&&\mbox{}
    +\left({\alpha_s(\mu)\over 4\pi}\right)^3
    \left(\tilde c_3^{[c]}(\mu r)+ \frac{64\pi^2}3 N_c^3\ln(\mu r)\right)
    +\cdots\Bigg]\,,
\end{eqnarray}
where
\begin{eqnarray*}
    \tilde c^{[c]}_1(\mu r) &=& \tilde a_1^{[c]} + 8\beta_0 \ln\left(\mu r e^\gamma\right)\,, \\
    \tilde c^{[c]}_2(\mu r) &=& \tilde a_2^{[c]} + 64\beta_0^2 \left[\ln^2\left(\mu r e^\gamma\right) + \frac{\pi^2}{12}\right] + \left(32\beta_1 + 16\beta_0 \tilde a_1^{[c]}\right) \ln\left(\mu r e^\gamma\right)\,, \\
    \tilde c^{[c]}_3(\mu r) &=& \tilde a_3^{[c]} + 512\beta_0^3 \left[\ln^3\left(\mu r e^\gamma\right) + \frac{\pi^2}4 \ln\left(\mu r e^\gamma\right) + 2\zeta(3)\right] \\
    &&\quad
    + \left(640 \beta_0\beta_1 + 192 \beta_0^2 \tilde a_1^{[c]}\right) \left[\ln^2\left(\mu r e^\gamma\right) + \frac{\pi^2}{12}\right] \\
    &&\quad
    + \left(128\beta_2 + 64\beta_1 \tilde a_1^{[c]} + 24 \beta_0 \tilde a_2^{[c]}\right) \ln\left(\mu r e^\gamma\right)\,,
\end{eqnarray*}
and
\begin{eqnarray*}
  \tilde a_1^{[c]} &=& a_1^{[c]}\,, \\
  \tilde a_2^{[c]} &=& a_2^{[c]}\,, \\
  \tilde a_3^{[c]} &=& a_3^{[c]} + \frac{64\pi^2}3 N_c^3 \gamma 
  \,.
\end{eqnarray*}

The corresponding relation in momentum space reads
\begin{eqnarray*}
  V^{[c]} &=&
  -{4\pi C^{[c]} \alpha_s(\mu)\over{\vec q}\,^2}
  \Bigg[1+{\alpha_s(\mu)\over 4\pi}c_1^{[c]}(\mu^2/\vec q\,^2)
    +\left({\alpha_s(\mu)\over 4\pi}\right)^2c_2^{[c]}(\mu^2/\vec q\,^2)
    \nonumber\\&&\mbox{}
    +\left({\alpha_s(\mu)\over 4\pi}\right)^3
    \left(c_3^{[c]}(\mu^2/\vec q\,^2) + 8\pi^2 N_c^3\ln{\mu^2\over{\vec q}\,^2}\right)
    +\cdots\Bigg]\,,
\end{eqnarray*}
where
\begin{eqnarray*}
	c_1^{[c]}(\mu^2/\vec q\,^2) &=& a_1^{[c]} + 4\beta_0 \ln\left(\frac{\mu^2}{\vec q\,^2}\right)\,, \\
	c_2^{[c]}(\mu^2/\vec q\,^2) &=& a_2^{[c]} + 16\beta_0^2 \ln^2\left(\frac{\mu^2}{\vec q\,^2}\right) + \left(16\beta_1 + 8\beta_0 a_1^{[c]}\right) \ln\left(\frac{\mu^2}{\vec q\,^2}\right)\,, \\
	c_3^{[c]}(\mu^2/\vec q\,^2) &=& a_3^{[c]} + 64\beta_0^3 \ln^3\left(\frac{\mu^2}{\vec q\,^2}\right) + \left(160 \beta_0\beta_1 + 48 \beta_0^2 a_1^{[c]}\right) \ln^2\left(\frac{\mu^2}{\vec q\,^2}\right) \\
  &&\quad
	+ \left(64\beta_2 + 32\beta_1 a_1^{[c]} + 12 \beta_0 a_2^{[c]}\right) \ln\left(\frac{\mu^2}{\vec q\,^2}\right)\,,
\end{eqnarray*}
and
\begin{eqnarray*}
  \beta_0 &=& \frac14\left[\frac{11}3 C_A - \frac43 T_F n_l\right]\,, \\
  \beta_1 &=& \frac1{16}\left[\frac{34}3 C_A^2 - 4C_F T_F n_l - \frac{20}3 C_A T_F n_l\right]\,, \\
  \beta_2 &=& \frac1{64}\bigg[\frac{2857}{54} C_A^3 - \frac{1415}{27} C_A^2
  T_F n_l - \frac{205}9 C_A C_F T_F n_l \bigg]\,.
\end{eqnarray*}


\section{\label{app::FR}Coordinate-space Feynman rules}

In coordinate space the QED Feynman rules for a static
lepton interacting with a photon read
\\[2em]
\begin{tabular}{ccc}
 \includegraphics{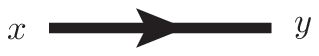} &
 $\theta\left(y_0-x_0\right)$ &
 \textbf{source propagator} \\ \\
 \includegraphics{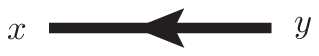} &
 $\theta\left(y_0-x_0\right)$ &
 \textbf{antisource propagator} \\ \\
 \includegraphics{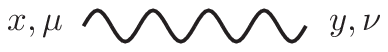} &
 $\frac{g_{\mu\nu}}{4\pi^2\left(y-x\right)^2}$ &
 \textbf{photon propagator} \\ \\
 \raisebox{-3em}{\includegraphics{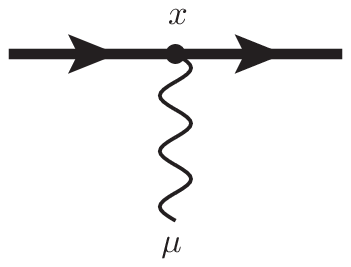}} &
 $-i g_s g^{\mu0} \delta\left(\vec x - \frac{\vec r}2\right) \theta\left(\frac{T^2}4 - x_0^2\right)$ &
 \textbf{source vertex} \\ \\ \\[-.5em]
 \raisebox{-3em}{\includegraphics{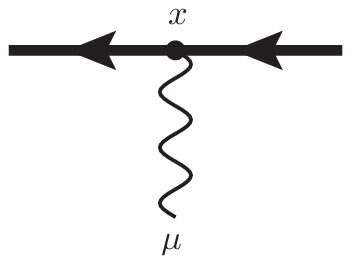}} &
 $i g_s g^{\mu0} \delta\left(\vec x + \frac{\vec r}2\right) \theta\left(\frac{T^2}4 - x_0^2\right)$ &
 \textbf{antisource vertex}
\end{tabular}
\\
The relation between the $g_s$ and $\alpha_s$ is given by 
$\alpha_s = g_s^2/(4\pi)$.


\end{appendix}



\end{document}